\documentclass{aa}  

\usepackage{graphicx}

\usepackage{txfonts}

\usepackage{natbib} 
\bibpunct{(}{)}{;}{a}{}{,} 

\usepackage{verbatim} 

\usepackage{hyperref}   

\hypersetup{colorlinks=true,linkcolor=blue,citecolor=blue,filecolor=blue,urlcolor=blue}

\usepackage{makeidx}
\makeindex

\begin{document}

   \title{Testing the asteroseismic estimates of stellar radii with surface brightness--colour relations and {\it Gaia} DR3 parallaxes}  

   \subtitle{III. Main-sequence stars}
\author{G. Valle \inst{1, 2}\orcid{0000-0003-3010-5252}, M. Dell'Omodarme \inst{1}\orcid{0000-0001-6317-7872}, P.G. Prada Moroni
        \inst{1,2}\orcid{0000-0001-9712-9916}, S. Degl'Innocenti \inst{1,2}\orcid{0000-0001-9666-6066} 
}
\titlerunning{KEYSTONE SBCR radii}
\authorrunning{Valle, G. et al.}

\institute{
        Dipartimento di Fisica "Enrico Fermi'',
        Universit\`a di Pisa, Largo Pontecorvo 3, I-56127, Pisa, Italy\\
        \email{valle@df.unipi.it}
        \and
        INFN,
        Sezione di Pisa, Largo Pontecorvo 3, I-56127, Pisa, Italy
}
   \date{Received 19/05/2025; accepted 13/06/2025}

  \abstract
{}
{
Expanding upon a recent investigation devoted to giant stars, we compare the radii derived from the asteroseismic scaling relations with those from surface brightness--colour relations (SBCRs) combined with the {\it Gaia} DR3 parallaxes for main-sequence (MS) stars.
}
{
The atmospheric and asteroseismic parameters were sourced from the recently released KEYSTONE catalogue and matched to \textit{Gaia} DR3 and TESS Input Catalog v8.2 to obtain precise parallaxes, $V$- and $K_S$-band magnitudes, and colour excesses, $E(B- V)$. We computed SBCR-based radii using two different SBCRs, and estimated their relative differences with respect to radius estimates from asteroseismic grid-based methods.
}
{
We find a good agreement between SBCR and asteroseismic radii, with mean relative differences in radii ($E_g$) in the range 2\% to 3\% and a standard deviation of about 3\%, consistent with the expected variability of SBCRs. We find no dependence on parallax, and a mild dependence on [Fe/H] for one of the SBCRs tested.
The relative difference in the estimated radii decreases as the mass increases, leading to a negative correlation between $E_g$ and the estimated stellar mass, with a slope varying from $-0.051\pm0.016$ to $-0.039\pm0.014$ per solar mass, depending on the chosen SBCR. This change in slope led to a roughly 1.5\% larger discrepancy in the $E_g$ estimates for stars with masses below 1.0 $M_{\sun}$. This larger discrepancy at the low-mass end supports conclusions drawn from giant star studies.
This result is independently corroborated by the LEGACY sample, which uses \textit{Kepler} photometry processed with the same pipeline as KEYSTONE. For the LEGACY sample we measure a mean relative offset in $E_g$ of $-1.4\%$ with a standard deviation of 2.3\%, and a dependence of $E_g$ on mass with a slope of $-0.052\pm0.011$ per mass unit, both fully consistent with the KEYSTONE analysis.
}
{The analysis reveals a strong agreement between SBCR-based and asteroseismic radii for MS stars, but the apparent mass dependence still requires closer examination. This result is reassuring as it demonstrates the great accuracy and reliability of the radius estimates obtained through SBCRs, which, moreover, offer the significant advantage of being applicable to a large sample of stars with substantially lower time and costs compared to what is required by asteroseismology. }

   \maketitle
   
\section{Introduction}\label{sec:intro}

Determinations of accurate stellar masses and radii are fundamental for constraining stellar structure and evolution models. While these parameters can be derived from the analysis of detached double-lined eclipsing binaries, obtaining them from single stars presents a more significant challenge.

Regarding stellar radii, few techniques allow astronomers to obtain precise estimates. 
One such method relies on surface brightness–colour relations (SBCRs). SBCRs provide an efficient way to determine stellar angular diameters from photometric measurements. Essentially, they establish a relationship between a star's angular size and its de-reddened brightness across various photometric bands. SBCRs are typically calibrated using samples of stars with well-determined radii \citep[see e.g.][]{Kervella2004,  DiBenedetto2005, Salsi2021, Kiman2024}.
Over 20 SBCRs exist in the literature, each focusing on different evolutionary phases. Many of these relations concentrate on the $V$ and $K$ bands, as this combination of colours offers the lowest dispersion \citep{Kervella2004}. When the distance to a particular star is known, applying an SBCR directly yields its linear radius.

A complementary technique for determining stellar radii has risen to prominence in recent years.  
Space-based asteroseismology — enabled by missions such as \textit{Kepler} and the Transiting Exoplanet Survey Satellite \citep[TESS;][]{Borucki2010,Ricker2015} — now provides independent estimates of fundamental stellar parameters, including mass and radius.
Neither the distance from the observer nor the reddening influence the asteroseismic estimates.
This method requires global seismic parameters, namely the frequency
of maximum power, $\nu_{\rm max}$, a large frequency separation, $\Delta \nu$, and a determination of classical atmospheric parameters, such as the stellar effective temperature and the metallicity, [Fe/H]. 
These parameters enable the estimation of stellar masses and radii either through classical scaling relations \citep{Ulrich1986, Kjeldsen1995} or via grid-based fitting. The grid-based approach generally delivers higher precision because it embeds stellar-evolution constraints by matching the observables to a pre-computed grid of evolutionary tracks. It is especially powerful for main-sequence (MS) stars and is now widely adopted in the literature \citep[see][]{Hekker2020}.

The aim of this paper is to compare the estimates of two techniques for MS stars. This work extends the previous investigations by \citet{Valle2024raggi}\defcitealias{Valle2024raggi}{Paper I} (hereafter \citetalias{Valle2024raggi}), and \citet{Valle2025raggi2}\defcitealias{Valle2025raggi2}{Paper II} (hereafter \citetalias{Valle2025raggi2}),
which focused on evolved red giant branch (RGB) and red clump (RC) stars. By employing the same methods and statistical techniques as in the earlier study, we aim to identify similarities and differences across various stellar evolutionary phases.
An ideal dataset for this comparison is the recently published KEYSTONE catalogue \citep{Lund2024}, which contains 173 stars with detected solar-like oscillations. The catalogue provides both atmospheric parameters (effective temperature, $T_{\rm eff}$, and metallicity, [Fe/H]) and global asteroseismic observables ($\Delta \nu$ and $\nu_{\rm max}$).

\section{Adopted SBCRs and data selection}\label{sec:data}

The stellar surface brightness in the $\lambda$ band,  $S_{\lambda}$, is linked to the stellar limb-darkened angular diameter, $\theta$, and its apparent magnitude
corrected from the extinction. In the $V$ band, $S_V$ is therefore
\begin{equation}
        S_V = V_0 + 5 \log \theta,   \label{eq:sv1}
\end{equation}
where $V_0$ is the $V$ band magnitude corrected for extinction. From Eq.~(\ref{eq:sv1}) it follows that
\begin{eqnarray}
        \theta &=& 10^{0.2 \; (S_V - V_0)}, \label{eq:theta}\\
        R_{\rm SBCR} &=& 0.5 \; d \; \theta, \label{eq:r}
\end{eqnarray}
where $d$ is the heliocentric distance of the star and $R_{\rm SBCR}$ its linear radius.

We adopted the \citet{Graczyk2021} SBCR, which links the limb-darkened angular diameter to the de-reddened magnitude,
\begin{equation}
        \log \theta = 0.2  \left(a_0 - V_0 + a_1  X + \ldots + a_5 × X^5 \right),
\end{equation}
where $X = (V-K_s)_0$ is the colour corrected for reddening. 
The fifth-order polynomial coefficients are $a_0 = 2.521$, $a_1 = 1.708$, $a_2 = -0.705$, $a_3 = 0.623$, $a_4 = -0.239$, and $a_5 = 0.0313$.
This relation, calibrated in a  $(V - K_s)_0$ colour range from $-0.2$ to 2.1 mag, has a mean precision of 1.1\% over the \citet{Graczyk2021} calibration sets consisting of 28 eclipsing binary systems with accurate radii determinations. This SBCR is consistent with the various dwarf star SBCRs in the literature within about 3\% \citep{Graczyk2021, Kiman2024}. The  \citet{Graczyk2021} SBCR in the $V$ and $K_s$ bands was chosen because it relies on the same photometric bands studied in   \citetalias{Valle2024raggi} and \citetalias{Valle2025raggi2}, allowing a direct comparison. 

To verify the robustness of the results, a different SBCR from \citet{Kiman2024} was also tested. This SBCR
leverages the more precise {\it Gaia} photometry, and was calibrated over 65 stars with known angular diameter.
We adopted the SBCR that relies on the $G$ magnitude and the $X = G-RP$ colour index:
\begin{equation}
\log S_G = \left( a X^2 + b X + c \right) \left( 1 + d \, {\rm [Fe/H]} \right),\label{eq:kim}
\end{equation}
with
$a= -0.09 \pm 0.03$, $b = 2.03 \pm 0.05$, $c = 0.85 \pm 0.02$, $d = -0.011 \pm 0.002$. This SBCR is valid in the colour range from 0.28 to 1.24 mag. 

Parallax values and their formal uncertainties were obtained through a nearest-neighbour cross-match between the KEYSTONE catalogue and {\it Gaia} Data Release 3 \citep[DR3;][]{Gaia2021}\footnote{All cross-matching in this work was performed with the Mikulski Archive for Space Telescopes (MAST) online tools.}.  
For every star the parallax signal-to-noise ratio exceeds 100, allowing distances to be safely estimated as the simple inverse parallax \citep{Bailer2021,Fouesneau2023}.
To obtain precise $V$ and $K_s$ band magnitudes and $E(B-V)$ for the objects in the KEYSTONE catalogue we cross-matched it with the TESS Input Catalogue (TIC) v8.2. The TIC adopts the 3D empirical dust maps from Panoramic Survey Telescope and Rapid Response System \citep{Green2018}, with a re-calibration coefficient of 0.884 applied to obtain  $E(B-V)$ values, as prescribed by \citet{Schlafly2011}.
We adopted the following extinction relations from \citet{Cardelli1989}: $A_V = 3.1 \, E(B-V)$ and $A_K = 0.114 \, A_V$. The adoption of the relation $A_K = 0.089 \, A_V$ from \citet{Nishiyama2009} does not modify the results, as well as the adoption of the extinction $A_0$ from \citet{Lallement2022} 3D dust maps.
{\it Gaia} magnitudes for the \citet{Kiman2024} SBCR were sourced from DR3 catalogue. Extinction in the broad {\it Gaia} $BP$ and $RP$ bands where computed from the \citet{Danielski2018} relations, using $A_0$ extinction, effective temperature, and log gravity.
 
Data in the merged catalogue were subjected to a selection procedure to exclude stars that had already evolved beyond the sub-giant branch into the early RGB phase and stars with unreliable astrometric solutions. To this aim,
only stars with $\log g > 3.8$, $T_{\rm eff} > 5500$ K, and \textit{Gaia} DR3 \texttt{RUWE} < 1.4 were retained.  The final sample comprises 87 stars. The selected stars had a mean metallicity [Fe/H] = $-0.07$ dex, with standard deviation $\sigma$ = 0.18 dex, mean parallax 11.4 mas ($\sigma$ = 4.0 mas), mean $(V-K_s)_0$ = 1.32 mag ($\sigma$ = 0.19 mag), and mean $(G-RP)_0$ = 0.43 mag ($\sigma$ = 0.04 mag).

\section{Comparing SBCR and asteroseismic radii }\label{sec:results-scale}

The KEYSTONE catalogue comprises asteroseismic data from three different pipelines: CV, SYD, and TACO/OCT. 
A total of 87 observations are available in the final dataset for the CV pipeline, 82 for the SYD pipeline, and 53 for the TACO/OCT pipeline. The analysis presented in the following adopts the CV data. Results from the other pipelines agree with the reported trends and suggest similar conclusions.
The adopted solar reference values were $T_{\rm eff,\sun} = 5777$ K, $\Delta \nu_{\sun} = 135.1$~$\mu$Hz and $\nu_{\rm max, \sun} = 3090$~$\mu$Hz from \citet{Huber2011}, which agree with the estimates reported by \citet{Viani2019} for CV pipeline.

\begin{figure}
        \centering
        \resizebox{\hsize}{!}{\includegraphics{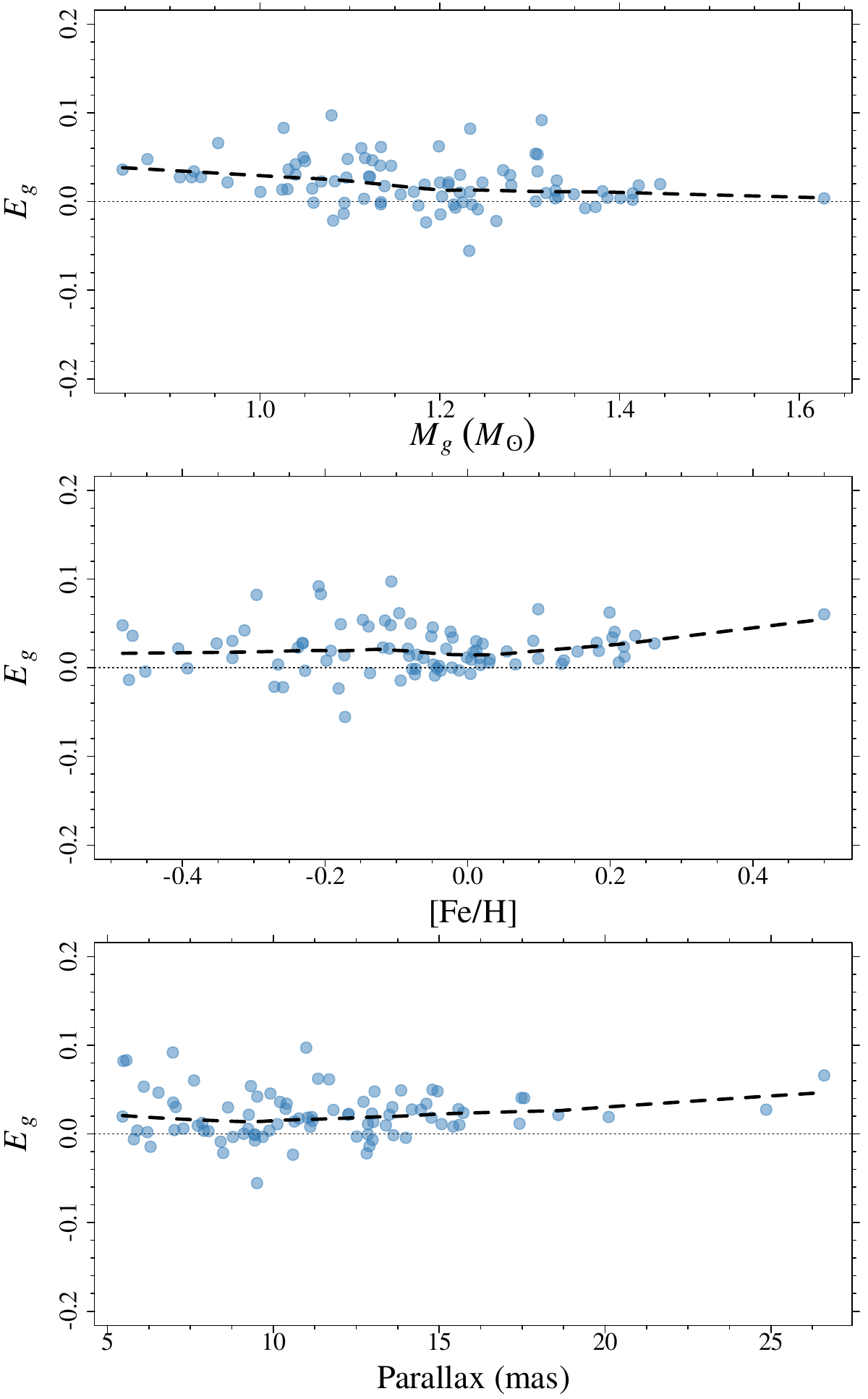}}
        \caption{Relative differences in the radii from the \citet{Graczyk2021} SBCR and the asteroseismic grid technique as a function of different parameters.
    {\it Top}: Dependence on the grid-estimated stellar masses. The dashed black line is a smoother of the data. {\it Middle}: Same as the top panel but as a function of the metallicity, [Fe/H].
    {\it Bottom}: Same as the top panel but as a function of the parallax value.}
        \label{fig:Eg-M}
\end{figure}

Asteroseismic grid-based estimates of stellar parameters were obtained 
using the SCEPtER pipeline\footnote{Publicly available on CRAN: \url{http://CRAN.R-project.org/package=SCEPtER}}, a well-tested technique \citep{eta}. The pipeline estimates the stellar parameters adopting a grid maximum likelihood  approach, relying on the observed quantities $o \equiv \{T_{\rm eff}, {\rm [Fe/H]}, \Delta \nu, \nu_{\rm max}\}$. 
The grid of stellar models used in this study is the same as that employed by \citet{Valle2024age} in their investigation on RGB star from APO-K2 catalogue \citep{Stasik2024}. 
Global asteroseismic parameters for grid models were obtained by means of the
 asymptotic scaling relations \citep{Ulrich1986, Kjeldsen1995}
\begin{eqnarray}
\Delta \nu &\propto& \sqrt{\frac{M}{R^3}}\\
\nu_{\rm max} &\propto& \frac{M}{R^2 \sqrt{T_{\rm eff}}}.
\end{eqnarray}

No correction was applied to the large–frequency separation, $\Delta \nu$, in our analysis.  
For near–solar–mass MS stars such adjustments are negligible; they become important only when reconciling observations with seismic–scaling estimates on the RGB \citep[e.g.][]{Epstein2014b,Gaulme2016,Viani2017,Rodrigues2017,Zinn2019,Stello2022,Li2023}.  
A direct test using the \citet{Li2023} prescription yields correction factors with a mean of 1.001 and a standard deviation of 0.018.  
Repeating the grid-based modelling with these corrected $\Delta \nu$ values produces differences that are well within the formal uncertainties, consistent with the findings of \citet{Valle2018}, who show that modest changes in $\Delta \nu$ have little influence on grid-based parameter estimates for MS stars.  
This choice is further validated a posteriori in Sect.~\ref{sec:legacy}, where we compare our results with grids adopting individual-frequency fitting.

\subsection{Comparison using the \citet{Graczyk2021} SBCR}

Asteroseismic grid-based estimates of radii and masses are denoted by $R_g$ and $M_g$, respectively.
Figure~\ref{fig:Eg-M} illustrates the trend in the relative differences between grid-based and SBCR radii.
 The plotted quantity $E_g$ is defined as
\begin{equation}
E_g = \frac{R_{\rm SBCR} - R_g}{R_{\rm SBCR}}.\label{eq:Eg}
\end{equation}
Asteroseismic grid-based radii exhibit a slight negative underestimation with respect to SBCR values, as the mean of $E_g$ is 2.1\% with a standard deviation of 2.7\%. This mean difference, which is influenced by the specific physics and chemical input used in the stellar model computations, is however within the 2\% to 3\% range of SBCR variability reported by \citet{Graczyk2021}. A comparison with \citetalias{Valle2024raggi} and \citetalias{Valle2025raggi2} reveals that the bias of about 2\% is equivalent to those found for RGB and RC stars by adopting K2 and {\it Kepler} datasets, while the dispersion for MS stars is significantly reduced, because standard deviations from 7\% to 10\% were found for giant stars. This bias reduction may stem from the fact that the MS sample comprises only nearby stars, in the 200 pc range, due to the intrinsic lower luminosity of MS stars, which limits the our ability to investigate them as a function of the distance from the observer. This is consistent with the results of \citetalias{Valle2024raggi} and \citetalias{Valle2025raggi2}, which show a strong reduction in the dispersion as a function of the parallax.

Figure~\ref{fig:Eg-M} shows there is no relation between $E_g$ and the parallax, and that there is a near-constant bias in the range covered by the data.
A robust linear model fit\footnote{
Robust linear models are an alternative to least-squares regression when the assumptions about the error distribution are violated. These models can downplay the influence of outliers and provide more accurate estimates in the presence of non-normal errors. Further details can be found, for example, in \citet{venables2002modern} and \citet{Feigelson2012}.}  has a slope of $0.001 \pm 0.001$ per mas, with the largest influence from the two stars with parallaxes greater than 24.  Analogously, no dependence was found on [Fe/H], with a robust linear slope of $0.014 \pm 0.015$ per dex.
There is, however, a dependence on the estimated mass. In fact, the mean bias is higher in the low mass end, and it reaches 3.6\% for $M \leq1.0$ $M_{\sun}$.  A robust linear model fit shows the presence of a dependence of $E_g$ on $M_g$ with a slope of $-0.051 \pm 0.016$ per mass unit. These findings are particularly interesting because 
a similar behaviours in the low mass end range were also reported in \citetalias{Valle2024raggi} and \citetalias{Valle2025raggi2} for RGB and RC stars. 
While the analysis of giant stars might be influenced by mass loss for RC stars and by the development of a relevant helium core, as discussed in 
\citetalias{Valle2024raggi} and \citetalias{Valle2025raggi2}, these factors do not influence MS stars in the explored mass range. 
The finding of a discrepancy, albeit minor, between SBCR and asteroseismic radii for $M \lesssim 1.0$ $M_{\sun}$ in a different evolutionary phase requires attention. 
Since Sect.~\ref{sec:legacy} shows that this discrepancy persists  when adopting different set of stellar models, different analysis techniques and a different sample of stars, it is possibly due to an intrinsic bias in one of the adopted techniques in this specific range of observables.
Further investigations may help to clarify the origin of this systematic discrepancy.

\subsection{Comparison using the \citet{Kiman2024} SBCR}

\begin{figure}
        \centering
        \resizebox{\hsize}{!}{\includegraphics{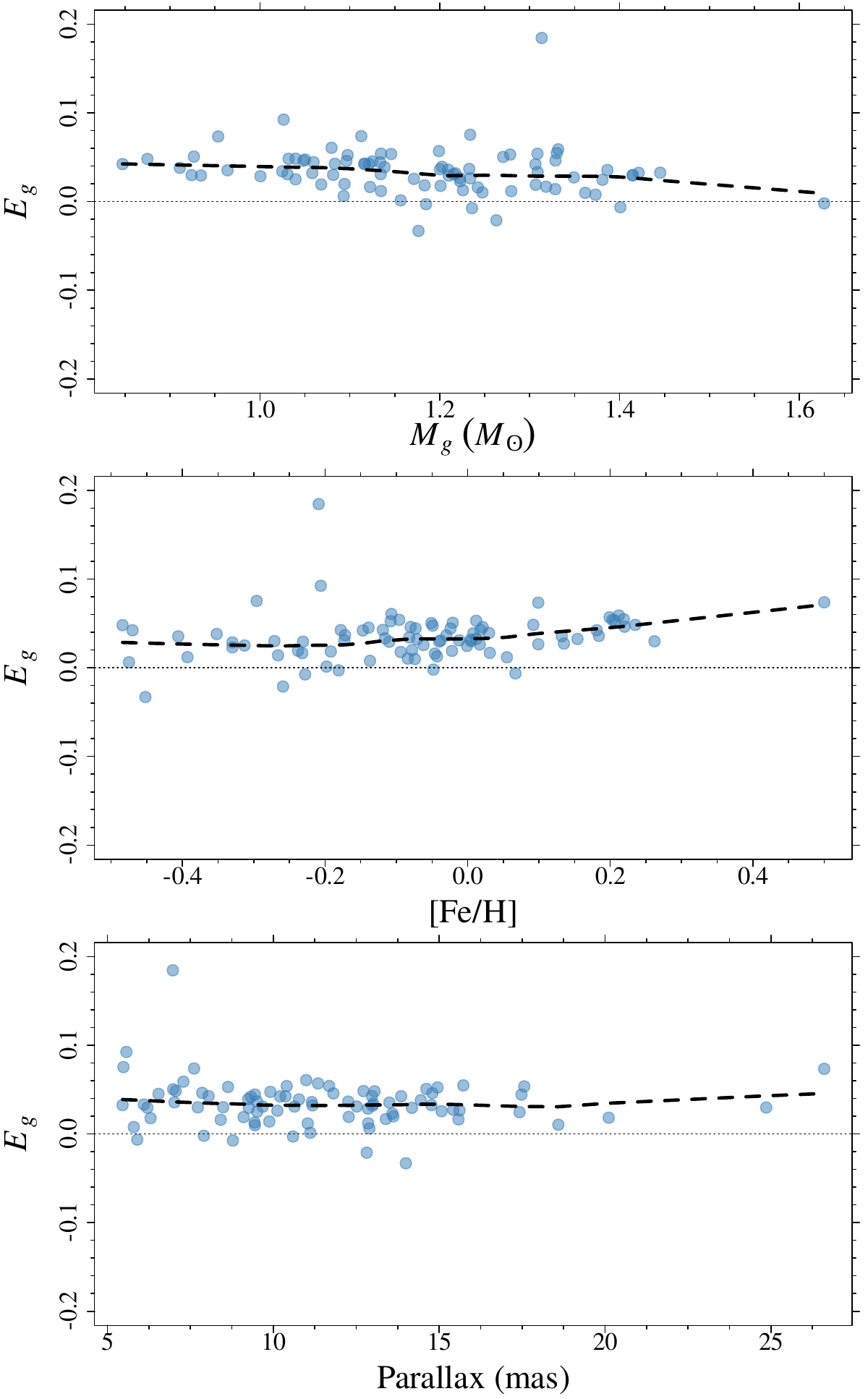}}
        \caption{Same as Fig.~\ref{fig:Eg-M} but using the \citet{Kiman2024} SBCR.}
        \label{fig:Eg-M-k}
\end{figure}

The analysis using the \citet{Kiman2024} SBCR shows a slightly higher discrepancy because SBCR radii are in mean 3.4\% higher than corresponding asteroseismic-based estimates, with a standard deviation of 2.6\%. The correspondence between the \citet{Graczyk2021} and \citet{Kiman2024} radii was noticeable, with a correlation coefficient of $\rho = 0.993$, with a 95\% confidence interval (0.990, 0.996). Given this agreement, it is not surprising that most of the results found for the \citet{Graczyk2021} SBCR still hold. In particular the presence of a trend in $E_g$ with respect to grid-based mass estimates is still present, even with a slightly reduced slope of $-0.039 \pm 0.014$ per mass unit. A dependence on the metallicity is present with a slope of $0.03 \pm0.01$ per dex. This detection is robust and persists even removing the highest metallicity star.
As for \citet{Graczyk2021}, no trend with the parallax was detected.

\section{A check with the LEGACY sample}\label{sec:legacy}

Since the asteroseismic grid-based estimates are inherently dependent on the computed stellar models, it is crucial to rule out these models as potential sources of the disagreement and trends reported in the previous section. While this hypothesis is highly unlikely, considering the overall high level of agreement between stellar models in the MS phase \citep[see e.g.][]{Stancliffe2016}, it must still be verified. 
In this section we compare our results for $E_g$ with those independently computed from the 66 LEGACY sample stars \citep{SilvaAguirre2017}. The LEGACY sample is a set of well-studied stars from the {\it Kepler} space telescope observations, and serves as a benchmark for asteroseismic analysis. By taking advantage of the high-precision data of the LEGACY sample, it was possible to estimate stellar masses and radii with a precision of about 1\% to 4\% in radii and 2\% to 4\% in mass \citep{SilvaAguirre2017}.

\citet{SilvaAguirre2017} provides masses and radii for the 66 stars, estimated using six different pipelines. These pipelines adopt different methods, stellar tracks computations, and observables. The majority of the pipelines adopted individual frequency fitting, computing theoretical frequencies by means of oscillation codes.
Given that the same observational data processing techniques were adopted for the KEYSTONE and LEGACY samples, this comparison is appropriate to investigate the presence of common trends over the two datasets.

\begin{figure}
        \centering
        \resizebox{\hsize}{!}{\includegraphics{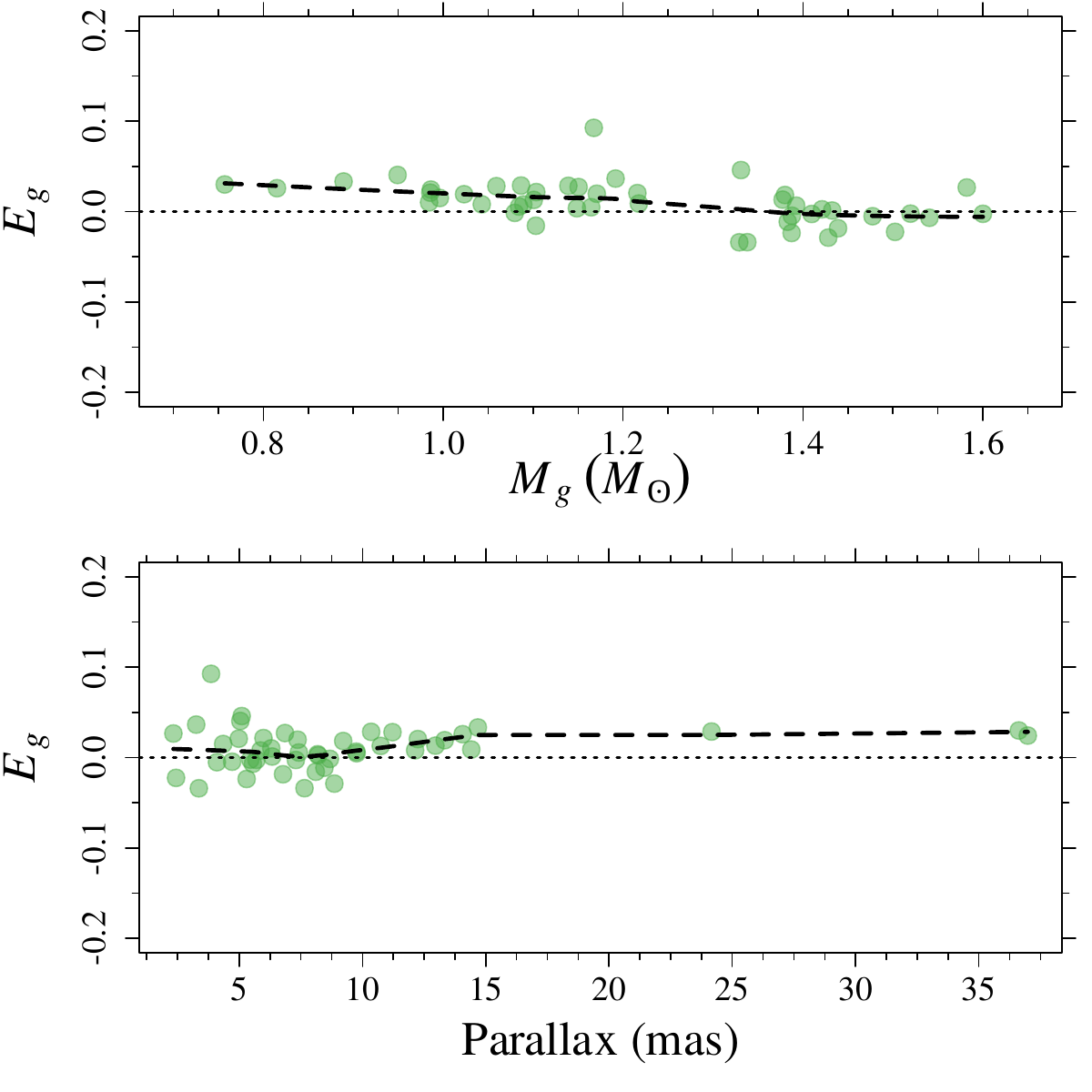}}
        \caption{Relative differences in the radii between a weighted average of the grid-based estimates from \citet{SilvaAguirre2017} for the LEGACY sample data and the \citet{Graczyk2021} SBCR estimates. {\it Top}: Relative differences as a function of the grid-based masses. {\it Bottom}: Same as the top panel but as a function of the parallax.}
        \label{fig:legacy}
\end{figure}

For each star in the LEGACY sample, we obtained grid-based mass and radius estimates using a weighted average of the values from the six pipelines in \citet{SilvaAguirre2017}. The $R_{\rm SBCR}$ values were then estimated on the LEGACY sample. Given the extremely high agreement between the \citet{Graczyk2021} and \citet{Kiman2024} SBCRs discussed in the previous section, we restricted the analysis only to the former SBCR.
To this aim, we cross-matched the LEGACY catalogue with \textit{Gaia} DR3 and TIC catalogues, using the same procedures outlined in Sect.~\ref{sec:data}. After excluding stars with insufficient astrometric quality, we obtained a final sample of 47 stars. 

As observed for the KEYSTONE sample, a trend of $E_g$ with estimated mass is present (top panel of Fig.~\ref{fig:legacy}).  The linear trend between $E_g$ and grid-based mass is $-0.052 \pm 0.011$ per mass unit, which closely aligns with the $-0.051 \pm 0.016$ per mass unit found for the KEYSTONE samples using our models grid.
Similar to the previous section, the grid-based radii are slightly underestimated compared to $R_{\rm SBCR}$, with a mean $E_g$ of 1.4\% (standard deviation 2.3\%), falling within the expected SBCR variability.
As with the KEYSTONE-based analysis, no dependences on [Fe/H] and parallax were found.
These results are expected given the agreement between asteroseismic grid-based estimates of radii among different grids in MS phase (see e.g. \citealt{SilvaAguirre2017} and  \citealt{smallsep} for a specific comparison of FRANEC-based estimates with those from different pipelines on the LEGACY sample).

\section{Conclusions}\label{sec:conclusions}

We compared radii derived from asteroseismic grid-based modelling  with those  from SBCRs combined with {\it Gaia} DR3 parallaxes, using the SBCRs  from \citet{Graczyk2021} and \citet{Kiman2024}. 
Expanding upon the work of \citetalias{Valle2024raggi} and \citetalias{Valle2025raggi2} for RGB stars, we investigated the MS phase using asteroseismic and atmospheric data from the recently released KEYSTONE catalogue \citep{Lund2024}. Information about colour excess and magnitudes in the $V$ and $K_s$ bands was obtained by cross-matching the KEYSTONE catalogue with TIC v8.2. 

The analysis revealed a very good agreement between the radius estimates: we find relative radius differences, $E_g$, of 2.1\% and a dispersion of 2.7\% for the \citet{Graczyk2021} SBCR, and 3.4\% with a standard deviation of 2.6\% for the \citet{Kim2002} SBCR, which are within the 2\% to 3\% range of SBCR variability reported by \citet{Graczyk2021}.
No dependence on parallax is observed, while a dependence on [Fe/H] is present when using the \citet{Kiman2024} SBCR.
A mass dependence emerges, with slopes of $-0.051 \pm 0.016$ and $-0.039 \pm 0.014$ per unit mass, when the SBCRs of \citet{Graczyk2021} and \citet{Kiman2024} are employed, respectively. These results reinforce the conclusions of \citetalias{Valle2024raggi} and 
\citetalias{Valle2025raggi2} for RGB stars and hint at an underlying bias in one of the radius‐estimation methods. 
It is important to note, however, that neither \citet{Salsi2021} nor \citet{Kiman2024} -- whose calibration samples cover the same mass range -- find any evidence of a systematic trend.

To verify that the observed trend is not an artefact of the adopted stellar models, we repeated the analysis with the model-based radii from the LEGACY catalogue \citep{SilvaAguirre2017}.  
The LEGACY radii were determined based on an independent grid of stellar models, a different inference pipeline, and a separate set of observables.  
With the \citet{Graczyk2021} SBCR, the mean relative offset between the KEYSTONE and \textsc{legacy} radii is $-1.4\%$ (standard deviation 2.3\%).
As in the KEYSTONE sample, we find that  $E_g$ is clearly dependant on mass, with a slope of  $-0.052 \pm 0.011$ per mass unit, in excellent agreement with the KEYSTONE slope.  
This concordance, despite the use of entirely different stellar models, strongly supports the robustness of the detection.

Overall, the analysis demonstrates a notable agreement between SBCR and asteroseismic grid-based radii. 
The dispersion of their relative differences is slightly higher for the KEYSTONE dataset with respect to {\it Kepler} based data, possibly due to the strong systematic inherent to K2 data, which was corrected by \citet{Lund2024} before their asteroseismic analysis.
However, the apparent dependence on stellar mass inferred from grid-based modelling warrants a closer investigation.

A comparison with the results of \citetalias{Valle2024raggi} and \citetalias{Valle2025raggi2} indicates that asteroseismic and SBCR radii agree far more closely on the MS than in later evolutionary phases, likely because the MS stars in our sample are located at much lesser distances than the giants. Given the considerable benefits of SBCRs in terms of observational time, instrumentation requirements, and financial constraints compared to asteroseismology, the findings presented in this series of papers suggest that SBCRs provide a way to obtain large datasets of asteroseismic-compatible radius estimates with minimal investment.

\begin{acknowledgements}
G.V., P.G.P.M. and S.D. acknowledge INFN (Iniziativa specifica TAsP) and support from PRIN MIUR2022 Progetto "CHRONOS" (PI: S. Cassisi) finanziato dall'Unione Europea - Next Generation EU.
\end{acknowledgements}

\bibliographystyle{aa}
\bibliography{biblio}

\end{document}